\documentclass[conference]{IEEEtran}

\usepackage{url}

\IEEEoverridecommandlockouts
\usepackage{cite}
\usepackage{amsmath,amssymb,amsfonts,bm}
\usepackage{algorithmic}
\usepackage{textcomp}
\usepackage{xcolor}
\def\BibTeX{{\rm B\kern-.05em{\sc i\kern-.025em b}\kern-.08em
    T\kern-.1667em\lower.7ex\hbox{E}\kern-.125emX}}

\usepackage{latexsym}
\usepackage{booktabs}
\usepackage[free-standing-units=true]{siunitx}

\usepackage{flushend}

\usepackage[pdftex]{graphicx} 
\DeclareGraphicsExtensions{.pdf}

\usepackage{paralist}	% Inline enumeration
\usepackage{enumitem}

\usepackage{xspace}

\usepackage{caption}
\usepackage{subcaption}
\newcommand{\eg}{e.g.,\xspace}
\newcommand{\ie}{i.e.,\xspace}

\newcommand{\etc}{etc.\xspace}

\newcommand{\rev}[1]{\textcolor{black}{#1}}
\newcommand{\Rev}[1]{\textcolor{black}{#1}}

\newcommand{\prob}{\mathrm{Pr}}
\newcommand{\params}{\bm{\theta}}

\newcommand{\setone}{\emph{Set1}\xspace}
\newcommand{\settwo}{\emph{Set2}\xspace}
\newcommand{\setthree}{\emph{Set3}\xspace}

\usepackage{hyperref}
\hypersetup{
    bookmarks=true,    	% show bookmarks bar?
    unicode=false,		% non-Latin characters in Acrobat's bookmarks
    colorlinks=true,	% false: boxed links; true: colored links
    linkcolor=black,	% color of internal links
    citecolor=black,	% color of links to bibliography
    filecolor=black,   	% color of file links
    urlcolor=black		% color of external links
}

\usepackage{fancyhdr}
\usepackage{acronym}

\acrodef{ac}[AC]{assurance case}

\acrodef{cdf}[CDF]{cumulative distribution function}
\acrodef{clec}[CLEC]{classification LEC}
\acrodef{cnn}[CNN]{convolutional neural network}
\acrodef{cte}[CTE]{cross track error}

\acrodef{darpa}[DARPA]{Defense Advanced Research Projects Agency}
\acrodef{dac}[DAC]{dynamic assurance case}
\acrodef{dnn}[DNN]{deep neural network}
\acrodef{dsc}[DSC]{dynamic safety case}

\acrodef{fcn}[FCN]{fully convolutional network}

\acrodef{gp}[GP]{Gaussian process}
\acrodefplural{gp}[GPs]{Gaussian processes} 
\acrodef{gpu}[GPU]{graphics processing unit}

\acrodef{iid}[i.i.d.]{independent and identically distributed}

\acrodef{lec}[LEC]{learning-enabled component}
\acrodef{les}[LES]{learning-enabled system}
\acrodef{lstm}[LSTM]{long short-term memory}

\acrodef{opav}[OPAV]{optionally piloted air vehicle}

\acrodef{pdf}[PDF]{probability density function}
\acrodef{pca}[PCA]{principal components analysis}

\acrodef{rl}[RL]{reinforcement learning}
\acrodef{rv}[RV]{random variable}
\acrodef{rlec}[RLEC]{regression LEC}
\acrodef{rnn}[RNN]{recurrent neural network}

\acrodef{uas}[UAS]{unmanned aircraft system} 
\acrodef{uq}[UQ]{uncertainty quantification}

\begin{document}    

\title{Towards Quantification of Assurance\\for Learning-enabled Components\\
\thanks{The \ac{darpa} has supported this work under contract FA8750-18-C-0094 
of the Assured Autonomy Program. The views, opinions, and/or findings 
expressed are those of the authors and should not be interpreted as 
representing the official views or policies of the Department of Defense 
or the U.S. Government.}}

\author{
	\IEEEauthorblockN{Erfan Asaadi, Ewen Denney, Ganesh Pai}
	\IEEEauthorblockA{\textit{SGT, Inc., NASA Research Park}\\
					Moffett Field, CA 94035, USA\\
					\{easaadi, edenney, gpai\}@sgt-inc.com}}

\maketitle              % typeset the title of the contribution

\acused{lec}

\begin{abstract}

\emph{Perception}, \emph{localization}, \emph{planning}, and \emph{control}, 
high-level functions often organized in a so-called \emph{pipeline}, are amongst 
the core building blocks of modern autonomous (ground, air, and underwater) 
vehicle architectures. 
These functions are increasingly being implemented using \emph{\aclp{lec}} 
(\acp{lec}), \ie (software) components leveraging knowledge acquisition 
and learning processes such as deep learning. 
Providing quantified component-level assurance as part of a wider (dynamic) 
assurance case can be useful in supporting both pre-operational approval of \acp{lec} (\eg by regulators), and runtime hazard mitigation, \eg using 
assurance-based failover configurations. 
%to manage error propagation between the functions 
%
%This paper develops a notion of assurance for \acp{lec}---based on identifying 
%and quantifying the relevant dependability attributes---discussing its relevance 
%and contribution to system-level assurance. 
%
This paper develops a notion of assurance for \acp{lec} based on 
\begin{inparaenum}[\itshape i\upshape)]
\item identifying the relevant dependability attributes, and 
\item quantifying those attributes and the associated uncertainty,  
using probabilistic techniques. 
\end{inparaenum}
%
%through a probabilistic treatment of uncertainty in the quantification 
%of those attributes.
%
%of the quantification of those attributes, along with their associated 
%uncertainty.
%
We give a practical grounding for our work using an example from the aviation
domain: an autonomous taxiing capability for an \ac{uas}, focusing on the application of \acp{lec} as sensors in the perception function. We identify 
the applicable quantitative measures of assurance, and characterize the 
associated uncertainty using a non-parametric Bayesian approach, namely 
Gaussian process regression. We additionally discuss the relevance and 
contribution of \ac{lec} assurance to system-level assurance, the generalizability
of our approach, and the associated challenges.  

%Using an aviation domain application---an autonomous taxiing capability for an 
%\ac{opav}---as an illustrative example, we specify and quantify \ac{lec} 
%attributes associated with assurance, also discussing their relevance and
%contribution to system-level assurance. 
%%
%\blurb{part of a larger concept of dynamic assurance case}

\end{abstract}


\thispagestyle{fancy} 

\section{Introduction}\label{s:intro}
%

%\subsection{Motivation}\label{ss:motivation}
%\blurb{Characterization of autonomy as being learning-enabled 
%(alternatively, increasing use of ML components towards enabling autonomy). }

Modern autonomous air, ground, and underwater vehicle architectures
use the high-level functions of \emph{perception} (converting sensed data into 
a model of the environment in which the vehicle is situated), \emph{localization} 
(establishing vehicle position and location within its environment), 
\emph{planning} (finding an optimal suite of actions to achieve goals such 
as following a path or trajectory, or navigation in general) and \emph{control} 
(regulating high-level vehicle behavior and low-level actuation as required), 
as core building blocks typically organized in a so-called 
\emph{pipeline}~\cite{lin2018}\rev{, \ie the functions are separate, broadly 
organized in the above order in a modular way, with the output of one 
function serving as the input to the next.} 

Increasingly, some or all of these functions are being implemented using 
\emph{\aclp{lec}} (\acp{lec}), \ie (software) components that leverage 
knowledge acquisition and machine learning processes. In particular, 
\acp{dnn} %~\cite{Goodfellow2016} 
have enjoyed a wide proliferation owing to the availability of 
large datasets and advances in \acp{gpu}~\cite{ai-dod}.

%\blurb{Autonomy assurance in safety-critical domains, e.g., aviation}
%\blurb{Key problems (cf. Google paper, Faria paper, other surveys)}

In safety-critical domains such as aviation, before systems can be deployed into civil airspace, assurance must be provided to various stakeholders 
(chiefly, the regulator) not only that those systems have been designed and 
developed to be inherently safe, but also that they can be operated at an 
acceptable level of safety. \rev{That, in turn, often also requires 
assurance that the constituent components perform adequately, are 
sufficiently reliable, \etc\footnote{\rev{In fact, components used on civil 
aircraft are required to meet prescribed minimum performance standards 
known as \emph{technical standard orders} (TSOs).}}}
As such, when \acp{lec} are being considered for use in aviation---\eg to 
realize transformative application concepts such as \emph{urban air 
mobility}~\cite{uam-landscape}---assurance of their fitness for purpose will 
inevitably be required both at a component and a system level, as appropriate. 
\rev{This is the primary motivation for the work in this paper, with the 
focus being on component-level assurance that \acp{lec} possess dependability 
attributes~\cite{dependability-taxonomy} additional to safety}.

%\subsection{Solution Context and Contributions}\label{ss:contributions}

\acused{ac} \acused{dsc} \acused{dac}

For novel aviation applications where regulations either do not 
yet exist, or are still being developed (\eg in enabling unmanned aircraft to 
fly beyond visual range), \emph{safety cases} have been a successful means by 
which to assure regulators of system safety~\cite{cdp-atio-2017}. 
\rev{An \emph{\acl{ac}} (\acs{ac}) is a generalization of the safety case 
concept meant to provide justified confidence that a system is \emph{fit for purpose}, and it addresses the broader system attributes, \eg of dependability.} 
The applicability of \acp{ac} is now being progressively explored for 
\acp{les}---\ie systems containing \acp{lec}---most notably for safety 
assurance of self-driving road vehicles~\cite{burton2017}. 

\rev{For the most part, \acp{ac} have used \emph{structured arguments}\footnote{A 
chain of reasoning that conveys the rationale why certain conclusions can be 
drawn (\eg of acceptable safety) based upon the evidence supplied.} as the 
mechanism to engender justified confidence}.
%
%for design-time assurance. 
%\cite{catapult2017}
%
Our prior work has advanced a notion of \emph{\acl{dsc}} 
(\ac{dsc})~\cite{dhp-icse2015}---more generally, \emph{\acl{dac}} 
(\acs{dac})---wherein a central tenet was \rev{using} \emph{confidence 
quantification}~\cite{dhp-esem-11} \rev{to provide assurance, in addition to structured arguments.}
\rev{The work in this paper, which we cast within this broader \ac{dac} framework, has two aspects that distinguish it from our prior work:} 
%
%within a framework for \emph{through-life safety}, \ie providing 
%assurance for the decision to deploy a system into operation, followed by 
%the continued provision and update of assurance both during and after operation. 
%%
%The \ac{dsc} concept subscribed to the fundamental principle that enabling
%continuous safety assurance requires assurance quantification, \eg in terms 
%of determining confidence in the claims made within a structured assurance 
%argument~\cite{dhp-esem-11}. In this paper, we reiterate that position, 
%asserting that assurance quantification is an essential component of a \ac{dsc} 
%and, more generally, a \emph{\acl{dac}} (\acs{dac}). 
%\blurb{Distinguish what is new in this paper, from the earlier DSC concept}
%
%Two aspects distinguish the work in this paper from this prior work: 
%
firstly, we focus \rev{on} component-level assurance quantification \rev{to be used}
within a wider \ac{dac}. Second, we 
%previously we used (the structure of) the 
%assurance argument for insight into the quantification model~\cite{dhp-esem-11}. 
%Here, we instead 
use arguments to relate system- and component-level 
dependability attributes~\cite{dependability-taxonomy}, and we apply \ac{uq} techniques focusing on the attributes of the \acp{lec}. 

%
%Moreover the argument structure gives insight into that which is being quantified
%and its relevance to other elements of the assurance argument. 
%
As its main contributions, this paper: 
%\begin{compactitem}[--]
\begin{inparaenum}[(i)]
	\item \rev{characterizes \ac{lec} assurance by} 
	identifying the relevant dependability attributes, specifying 
	\emph{assurance properties}, and developing a notion of \emph{assurance 
	measure} (\rev{see Section~\ref{ss:terminology}}); and %
	%(Section~\ref{sss:characterizing-assurance}); 
	%
%	\item discuss the relevance and contribution of component-level assurance
%	to system-level assurance within the framework of our notion %\footnote{
%%	In contrast to that developed in~\cite{calinescu2017}.} 
%	of \acp{dac}. 
	%
	\item \rev{provides} a practical grounding to illustrate 
	\rev{the} feasibility of assurance quantification using 
%	the above using 
	an aviation system example: an autonomous taxiing capability 
	for an \acf{uas}, focusing on the application of \acp{lec} as sensors in 
	the perception function. Specifically, we elaborate one solution to 
	quantify \ac{lec} assurance, based on Bayesian non-parametric \ac{uq} 
	techniques, in particular \ac{gp} regression. %(Section~\ref{s:example}).  
\end{inparaenum}
%\end{compactitem}

We additionally discuss the relevance and contribution of component-level 
assurance to system-level assurance within a wider framework of \acp{dac}.

%\blurb{probabilistic model checking, runtime verification}

%\subsection{Dynamic Assurance}\label{ss:dynamic-assurance}

%\blurb{Brief background on DACs, with AQM being part of the DAC component that is relevant for this paper}
%
%
%\blurb{Elaborating on the \emph{dynamic} bit - updates based on \emph{conditional
%evidence}}
%
%\blurb{Updates to various components, including arguments. Explain notion of conditional evidence.}
%
%\blurb{Main focus in this paper is on assurance measures and providing a rigorous basis for that notion}
%
%\blurb{Explain how component assurance measures and their quantification as in the
%preceding discussion fits into a DAC} 

\section{\rev{Related Work}}\label{s:related-work}

%\blurb{Previous work on assurance quantification}

Our prior work on \ac{ac} confidence quantification~\cite{dhp-esem-11}
is closely related to the work in this paper. %As mentioned earlier (Section~\ref{s:intro}), 
Where that work sought to build the quantification model based on the 
structure of the underlying argument, the current work proposes to use arguments
mainly for the decomposition and refinement of assurance properties. Also, for assurance quantification we use Bayesian non-parametric \ac{uq} techniques. 
%
%relies on Bayesian non-parametric \ac{uq} techniques instead. However, the 
%approach we take to developing \ac{lec} assurance properties and measures is 
%closely tied to the methodology used to develop assurance arguments. 
%
Other literature on confidence quantification for assurance purposes, \eg \cite{guiochet2019}, has investigated the application of evidential theory, 
although it has not been applied to the assurance of \acp{lec}.

%\blurb{Dynamic safety cases} 

Our original concept of \acf{dsc}~\cite{dhp-icse2015}, \rev{in which assurance 
quantification was a core principle}, has since been applied for engendering 
trust in adaptive software~\cite{calinescu2017}. \rev{Although that work 
has not considered \acp{lec}, it} leverages probabilistic model-checking 
techniques, which can provide a quantitative notion of assurance 
\rev{(akin to what we will develop later, in Section~\ref{s:example})}, 
if \acp{lec} can be represented using the state-space models \rev{that are} 
compatible with the verification techniques, \eg as in~\cite{verisig}.
%
%\eg as in~\cite{reluplex} and~\cite{verisig}. 
%
Nevertheless, the core aspect of assurance \rev{in} that work is an 
updatable assurance argument, rather than a quantified assurance measure.

Analogous to our notion of \acp{dsc}, dynamic safety management~\cite{trapp2018} 
is a \rev{proposed} run-time methodology for assurance \rev{given in terms of 
quantified} safety risk. However, this plausible theoretical basis for 
quantification is largely speculative about the applicability to \acp{lec}.

%
%This work focuses on operational updates of the associated structured 
%argument(s) with evidence from runtime verification. 
%%
%Usually since a human (operator, safety manager, or regulator, say) is the 
%main consumer of such arguments (and the corresponding assurance case), at 
%runtime the \ac{les} itself (or, alternatively, the \acp{lec}) may have limited 
%use for dynamically updated arguments without associated machine processable 
%semantics\footnote{The proposal to align structured assurance 
%arguments with `natural language deductivism'~\cite{driscoll2018} is 
%a plausible approach towards such semantics.}, computation, or metrics 
%enabling continued and assured functioning. 
%%
%To our knowledge, these executable capabilities are not (yet) facets of 
%structured arguments employed in production safety cases. %do not yet 
%%provide any of these.
%%
%
%It has been suggested~\cite{weaver-bowtie-2012}, albeit in the context of 
%non-\acp{les}, that operational assurance may be better conveyed through 
%mechanisms that reflect how risk changes from an accepted 
%and approved baseline\footnote{Here, regulatory approval and acceptance of the risk 
%baseline may be obtained using argument-based assurance cases.}.
%%
%We have successfully confirmed this in practice, and extended the underlying 
%models to capture the broader scope of safety whilst providing a basis for 
%quantifying the risk reduction achieved, one of the metrics for 
%assurance~\cite{dpw-safecomp-2017}.

Quantification of application independent \ac{lec} (assurance) properties has 
also been explored in the recent literature, \eg using Bayesian neural networks 
(NNs) to quantify model uncertainty~\cite{gal2018}, and using \acp{gp} to 
quantify robustness against adversarial examples and input 
perturbations~\cite{Kwiatkowska2019}. \rev{This is compatible with our 
methodology (Section~\ref{ss:methodology}), although what differentiates our work 
is that our approach links such quantified \ac{lec} properties to component
and system-level dependability attributes, thereby giving a domain-specific 
semantics to quantification, in terms of assurance}.

\section{\rev{Solution Overview}}\label{s:approach}

\subsection{\rev{Concepts and} Related Terminology}\label{ss:terminology}

\begin{figure}
	\centering
	\includegraphics[width=\columnwidth]{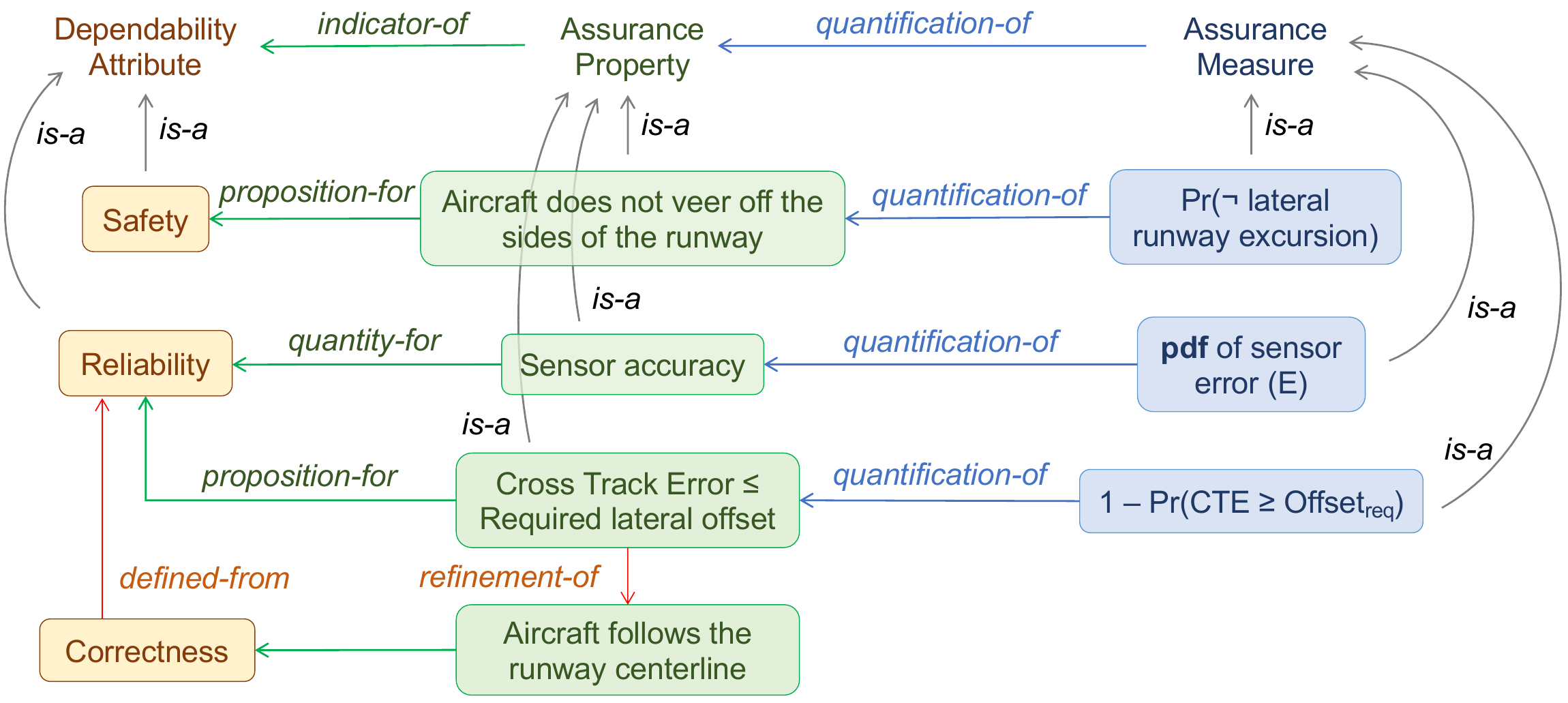}
	\caption{Assurance properties and assurance measures: concepts facilitating 
	the characterization of assurance. \rev{The properties and measures shown relate
	to our illustrative example (Section~\ref{s:example}).}}
	\label{f:assurance-measures}
\end{figure}

%First we introduce the terminology related to our proposed methodology, 
%referring to  Fig.~\ref{f:assurance-measures}, which illustrates the 
%associated concepts. 

%\begin{asparaenum}[\itshape a\upshape)]
%\item
\emph{Assurance} is (the provision of) \emph{justified confidence} that a system 
\rev{(or component/item, or service)} possesses the applicable dependability attributes. 

\acused{cte}

An \emph{assurance property} is a (logical or probabilistic) proposition, or 
a quantity, associated with a dependability attribute. 
For instance (Fig.~\ref{f:assurance-measures}), \rev{sensor accuracy} is a 
component-level assurance property that is also a quantity associated with
the dependability attribute of \emph{reliability}. \rev{On the other hand, the 
proposition that (aircraft) \emph{\acl{cte}} (\acs{cte}) is within the bounds of a 
required lateral offset is a \emph{system-level} reliability assurance property. 
That, in turn, is a refinement of a \emph{correctness}\footnote{\Rev{We include correctness as an attribute to admit a formal methods concept for software, and relate it to the more general dependability attribute of reliability.}} 
assurance property, \ie 
the proposition that the aircraft follows the runway centerline. 
\Rev{Specifically, showing (over time) that the system correctly maintains 
\acs{cte} within the required lateral offset bounds provides assurance for 
reliability, refining the required correct behavior that the aircraft follows 
the runway centerline.}
Likewise, the proposition that the aircraft does not veer off the sides of a
runway is a system-level \emph{safety} assurance property.}

%
%that is an indicator of the dependability property of safety}. 
%
%the time to sensor failure is a 
%component-level assurance property that is also a quantity associated with 
%the attribute of reliability. The proposition that actuator reliability 
%meets the required reliability is, likewise, a component-level assurance 
%property indicative of the attribute of reliability.
%

%\item
An \emph{assurance measure} is a suitable probabilistic quantification of an 
assurance property---\ie both the interpretation and the associated numerical 
values that are assigned to the underlying quantity or proposition. 
Candidate assurance measures corresponding to the previously mentioned 
(reliability) assurance properties (Fig.~\ref{f:assurance-measures}) are, respectively, the \rev{\ac{pdf} of the sensor error, and the probability that 
the \ac{cte} does not exceed the required lateral offset.}
%
%the \ac{cdf} of the survival time, and 
%the probability that the actuator reliability requirement is met.
%For the latter, note that we consider actuator reliability itself as a 
%\ac{rv} to account for \emph{uncertainty}. 
%
%\item
The \emph{uncertainty} in an assurance measure expresses the \emph{confidence} 
associated with the assurance property, and thereby quantifies the extent of assurance. 	
%\end{asparaenum}

In general, we can develop a plurality of assurance properties and 
assurance measures for a given dependability attribute. 
Attributes such as safety are largely meaningful in a system context, although
\rev{we can formulate} assurance properties (both at a system- and a 
component-level) \rev{that are related to other dependability attributes, and 
that have a bearing on safety. For example, when a component is part of a safety 
system, its reliability (correctness) directly contributes to the delivery 
of the safety function and, thereby, to system safety}. 

\rev{From Fig.~\ref{f:assurance-measures} we can intuit that---based on 
the dependability attributes being considered---assurance properties 
correspond to the subset of both functional and non-functional requirements that 
can be reasonably quantified via assurance measures. As we will see subsequently (Section~\ref{ss:methodology}), assurance properties for a system and its constituent components are themselves interrelated.} 
Next, we characterize the scope of \ac{lec} assurance as context for our approach
to quantification (and its generalizability).

%
%\blurb{Explain the relationship between F/NF requirements and assurance 
%properties - assurance properties are the subset of requirements that are 
%quantifiable} 
%

\subsection{Problem Scope}\label{ss:scope}

%\blurb{Scope of the general problem as 
%
%Function X Type of Learning X Learning problem X Learning scheme X network architecture X assurance concerns
%
%For example
%
%Perception X Supervised X Regression X Deep learning X CNN X Dataset shift or  covariate shift; example use as sensors to predict various continuous quantities
%such as distance, speed, temperature, heading, etc. 
%
%Perception X supervised X classification X deep learning x CNN X Dataset shift: 
%example use as sensors for object detection, tracking
%
%Localization X supervised X classification x SVM X ...
%}

There are number of choices to make when using \acp{lec} to implement the high 
level functions of perception, localization, planning, and control, in an 
autonomous vehicle architecture. 
For instance, we can either implement each function in the pipeline, individually, as an \ac{lec}; alternatively, a combination of functions (\eg perception and 
localization) may be candidates for \ac{lec}-based implementation. There are 
also so-called \emph{end-to-end} implementations in which a single \ac{lec} 
maps sensor inputs to low-level control outputs~\cite{bojarski2016}.

Next, an extremely rich variety of learning schemes, algorithms, and models are 
available to construct \acp{lec} suitable to the requirements, \eg perception 
and localization functions have been implemented as (deep) \acp{cnn} trained 
using supervised learning. Based on the sensors available, 
%(such as multi-spectral cameras, radars, sonars, lidars, \etc) 
such deep \acp{cnn} can be used not only for detection, classification, and 
tracking of various entities in the operating environment (broadly, 
classification problems), but also for predicting continuous quantities such 
as distance, speed, heading, \etc (more generally, regression problems). 
Deep \acp{cnn} have also been shown to be applicable for implementing the 
control function~\cite{bojarski2016}, while alternative implementations for 
planning and control functionality can use \acp{dnn} (which may or may not 
include convolutional layers) trained using \ac{rl}~\cite{dnn-rl}.
%, \eg as in~\cite{dnn-rl}.  

The preceding options represent only a small selection from the space of 
possible \ac{lec} implementation choices, and the literature is replete with 
other novel combinations of learning models and schemes~\cite{xu2017}. 
As such, \acp{lec} pose particular assurance challenges that are, largely, 
specific to the algorithms used, implementation choices made, and their 
application and usage context. For instance, \ac{rl} poses the risk of 
\emph{reward hacking}: exhibiting unexpected emergent behavior wherein a 
policy is learned that attains the reward objectives at the expense of 
the (longer term) mission or safety objectives~\cite{amodei2016}. 
%\cite{faria2018}. 

\acp{lec} %developed through supervised learning as well as \ac{rl} 
are susceptible to \emph{dataset shift}, \ie encountering environments and data not representative of those used for training.
%\footnote{\acp{lec} developed using \ac{rl} are also affected, although the 
%effects are amplified~\cite{amodei2016}.} 
%
%\blurb{Need for assurance quantification - objective decision support 
%in a safety and mission critical context}
%
%Moreover, since \acp{lec} %(\ie their outputs or behaviors) 
%are expected to generalize to situations on which they have not been trained, 
%the errors produced\footnote{\emph{Error} in this context refers to the difference
%between the \ac{lec} prediction (its output), and the true or required value. This 
%error may or may not be an \ac{lec} \emph{failure}, \ie when the error exceeds 
%a specified bound.} may manifest as safety-critical 
%consequences~\cite{tesla-nhtsa}, when propagated through the pipeline from 
%sensing to actuation (usually in combination with other system states and 
%environmental conditions). 
%
%
Moreover, since \acp{lec} %(\ie their outputs or behaviors) 
are expected to generalize to situations on which they need not have been trained, 
the errors produced may manifest as safety-critical 
consequences~\cite{tesla-nhtsa}, when propagated through the pipeline from 
sensing to actuation (usually in combination with other system states and 
environmental conditions). Note that, in this context, \emph{error} refers to 
the \rev{discrepancy} between the \ac{lec} prediction (its output), and the
rev{required or expected output}. This error may or may not \rev{lead to} 
a \ac{lec} \emph{failure}, \eg when the error exceeds a specified bound.

%\blurb{Discuss the specificity/generality of our approach and solution}

%In its current form, our methodology is primarily applicable to \acp{lec} 
%that have been trained in an offline supervised learning scheme, for 
%application to regression and classification problems, and where there is a 
%notion of true reference values against which \ac{lec} output can be compared. 

%\blurb{could be moved to discussion/future work section,
%indicating future work as extending to \ac{rl} problems}.

Next, we propose a generalized methodology for \ac{lec} assurance quantification; although, in light of the preceding discussion, we anticipate that concrete 
solutions---instances of our methodology---for providing (and quantifying) 
assurance are likely to be specialized to the particular application or usage 
of \acp{lec}, and the implementation choices made.

\subsection{Methodology}\label{ss:methodology}

Our overall approach to \ac{lec} assurance quantification involves: 
\begin{inparaenum}[1)]
	\item characterizing assurance in terms of the dependability attributes, 
	assurance properties, and assurance measures, taking into account the 
	application/usage of the \acp{lec} and their implementation; 
	and 
	\item applying \acf{uq} techniques to determine the numerical values to 
	be assigned to the identified assurance measures.
\end{inparaenum}

\subsubsection{Characterizing Assurance for \acp{lec}}
\label{sss:characterizing-assurance}

Relating assurance to dependability attributes (as we have in the 
preceding narrative) broadly encompasses the compatible notion of assurance 
as the confidence that a system functions as intended in its usage environment. 
Here, two insights are noteworthy: first, \rev{as mentioned earlier 
(Section~\ref{ss:terminology})}, assurance properties can be seen as 
\rev{a subset of the} appropriate system- and component-level 
requirements \rev{(in particular, those that can be reasonably quantified)}, so 
that the evidence that the requirements are satisfied contributes to increased 
assurance. This lends itself to the application of the entire suite of 
techniques for requirements decomposition and refinement, \eg as in~\cite{kaos2001}, 
\rev{in order to develop assurance properties and measures for a given \ac{lec}.} 
%
%\blurb{In general, typically the case that not all requirements would be 
%satisfied in an operational system. Here we do not discuss prioritization or 
%trade-offs between conflicting requirements}
%
%
%\blurb{Explain what is new and what we have developed} 
%\blurb{Explain system vs. component assurance} 
%
%\blurb{Relate above two to motivation} 
%
%
%\blurb{Explain how this fits in within the wider context of (dynamic) assurance
%cases - lends to idea of qualitative assurance versus quantitative assurance}
%
%

Second, we can map assurance properties to the claims made in an assurance 
argument, wherein a combination of inductive and deductive reasoning linking 
those claims to the evidence supplied, provides the basis to conclude that 
the claims have been met. 
\rev{For instance, in Fig.~\ref{f:assurance-measures}, we can state 
the system-level reliability and correctness assurance properties as 
claims in a structured argument that 
\begin{inparaenum}[(i)]
	\item provides the rationale for the \emph{refinement-of} relation between
	the respective assurance properties;  and 
\item further refines/decomposes the reliability assurance property into 
	lower-level assurance properties, \eg of sensor accuracy, that can be 
	ultimately linked to the substantiating evidence items.
\end{inparaenum}
Such arguments in themselves provide qualitative assurance, supplementing 
that obtained from quantifying the associated assurance measures.}
Thus, another approach for developing assurance properties and measures for 
a given \ac{lec} application context and is to create structured assurance arguments, for example, using the methodology in~\cite{dp-jase2018}. 
%
%In brief, this involves a combination of both 
%\emph{top-down} reasoning (\ie given the system-level dependability attributes, 
%determine the relevant \ac{lec} assurance properties and measures) and 
%\emph{bottom-up} reasoning (\ie given the available metrics, determine their 
%contribution to component- and system-level assurance) to \emph{meet in the 
%middle}. 
%
Note that the two approaches that we have mentioned here are themselves
complementary. 

\subsubsection{Uncertainty Quantification}\label{sss:uq}

%\blurb{Describe what we are doing w.r.t. quantification to highlight what 
%we are doing in this paper} 

We can apply Bayesian parametric or non-parametric \ac{uq} techniques for 
assurance quantification, based upon the specific assurance measures being considered. 
%
%\blurb{Parametric approach}
%
%We present the main idea underlying the Bayesian parametric approach using 
%an example of an \ac{lec} performing regression, where an assurance property 
%can be, say, reduced \emph{regression error}, $E$ and the corresponding measure 
%is an estimate of the expected error together with the associated uncertainty.
%
%The idea in the parametric approach is: firstly, we model the assurance measure
%as a \acf{rv}, say $X$, that is assumed to follow a known distribution from a 
%model class $M_i$, with model parameters $\bm{\theta}$, \ie 
%$X \sim \prob(X\mid\bm{\theta},M_i)$
%
\paragraph{Bayesian Parametric \ac{uq}}
Here, a parametric probability distribution is assumed to represent the 
uncertainty associated with the \ac{lec} assurance measure of interest. 
If $A$ is a \acf{rv} for the assurance measure, $\params$ are distribution 
parameters, and $M_i$ represents a class of probability distributions 
selected from a set of available candidate distributions, then the 
probability distribution $A \sim \prob(A | \params, M_i)$ characterizes 
the associated uncertainty. 
The idea, then, is to identify the parameters of the probability distribution 
in a Bayesian framework.
To this end, the prior probability distribution of the parameters, 
$\prob(\params | M_i)$ (representing our prior knowledge), is weighted 
by the likelihood of the parameters, 
$\prob\left(\mathbf{E} | \params, M_i \right)$, based on observed
\emph{evidence}, $\mathbf{E}$. %\eg \blurb{generic example of evidence}. 
Normalizing the weighted prior using Bayes' theorem, we can determine the 
posterior distribution of the parameters, 
$\prob \left(\params | \mathbf{E}, M_i \right)$, to establish not only
the expected value of the parameters but also the associated uncertainty (\eg 
in the model identification process). For more details, we refer the reader
to~\cite{asaadi2017computational}.

%For more detail about model parameter identification and model class selection 
%see [1-2] and for the design of the training set see [3]

\paragraph{Bayesian Non-parametric \ac{uq}}

Where parametric \ac{uq} techniques are appropriate for a limited number of 
parameters, Bayesian non-parametric models relax the assumption of a particular
parametric form for the uncertainty distribution. 
That is, rather than assuming a distribution with random parameters, 
the \ac{rv} for the assurance measure is a random function over which a prior 
is assumed, and updated within a Bayesian framework.   
Effectively, a much larger (potentially infinite) parameter space is admitted, 
with greater flexibility in representing the nature of uncertainty, although 
there is a trade-off with respect to computational cost.

A \acf{gp}~\cite{gp-book} is a versatile prior, often used in non-parametric Bayesian \ac{uq}, representing a continuous stochastic process such that the 
joint distribution of the functions over which it is a prior, is (assumed to be)
a multivariate Gaussian. A \emph{covariance} function is a central concept 
to \acp{gp}, and it models the correlations between different points in the 
stochastic process. 
We adopt non-parametric Bayesian \ac{uq} using \acp{gp} as our approach to \ac{lec}
assurance quantification and defer further discussion on the specifics to  
Section~\ref{ss:lec-assurance-q}, where we illustrate its application in the 
context of a running example (discussed next).

\section{Illustrative Example}\label{s:example}

We now present an example drawn from the aviation domain to illustrate 
the application of our methodology for \ac{lec} assurance quantification. From 
the standpoint of \acp{lec}, the example highlights their usage as sensors in the 
perception function. More specifically, we consider deep \acp{cnn} trained 
offline in a supervised learning scheme for regression problems. 

\subsection{System Description}\label{ss:system-overview}

The target application is an autonomous taxiing capability to be deployed in 
an \acf{uas}. The overall goal is to facilitate the aircraft taxiing on an 
airport taxiway or runway without human pilot input, whilst meeting the 
applicable functional and safety objectives (described next in 
Section~\ref{ss:system-level-assurance-scope}). To meet this goal, the 
perception function ingests video from a wing-mounted camera pointed towards 
the nose of the aircraft, producing output that is then processed by the 
localization, planning, and control functions, which, in turn, actuate and move 
the aircraft as required. 

\begin{figure}
	\centering
	\includegraphics[width=\columnwidth]{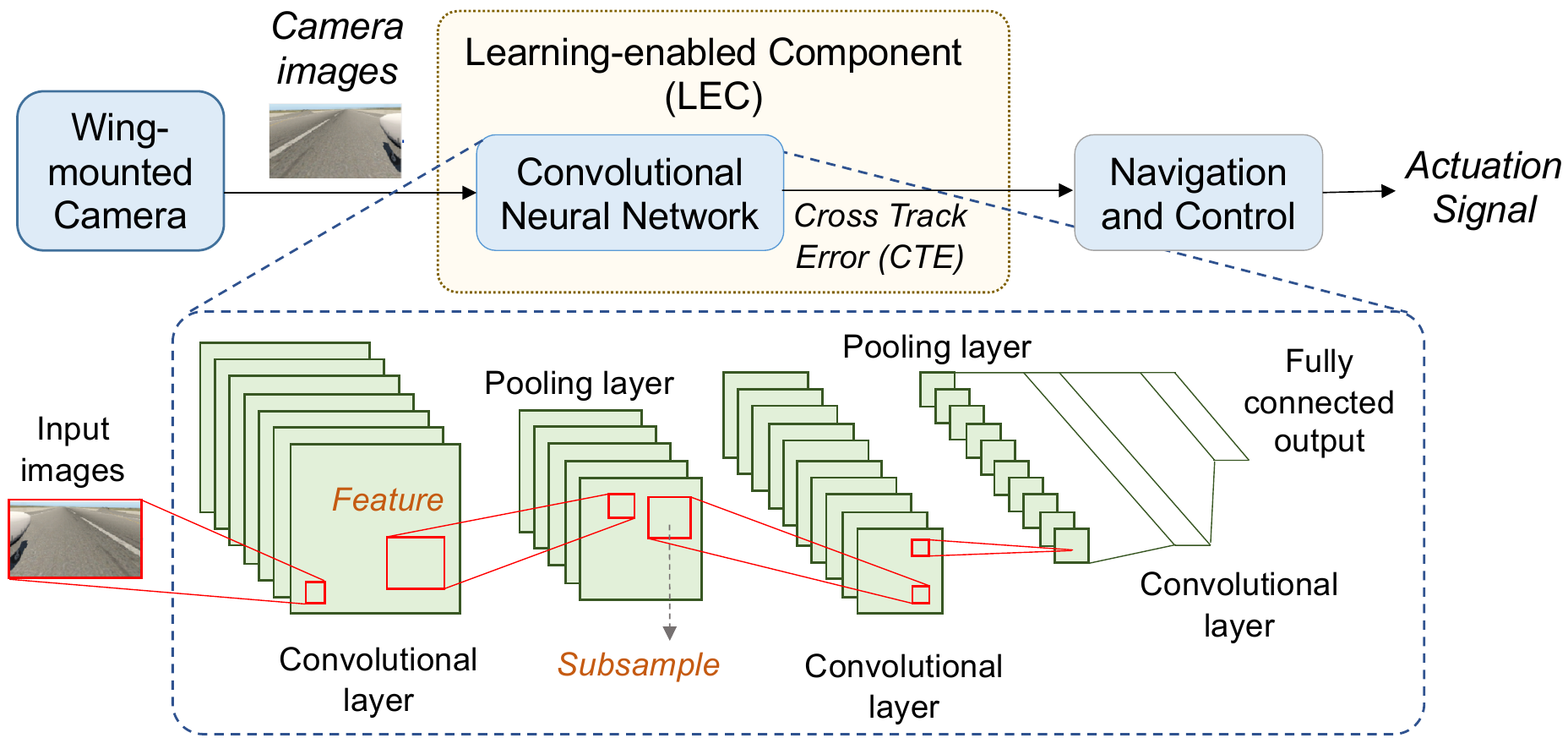}
	\caption{Notional deep \ac{cnn} implementing an \ac{lec} 
	that %functions as a (distance) measurement sensor, 
	processes camera images and estimates \acf{cte}, as part of the perception 
	function for an autonomous taxiing capability in an \acl{uas}.}
	\label{f:lec-example}
\end{figure}

Fig.~\ref{f:lec-example} shows a simplification of this pipeline in which the 
perception function is an \ac{lec} that is implemented, notionally, as a deep 
\ac{cnn} processing camera images as input. The input layer is $(360 \times 200)$ 
pixels $\times~3$ channels wide; the network size and complexity is of 
the order of $100$ layers with greater than two million tunable parameters. 
The \ac{lec} output of interest for this paper is a prediction of the 
\emph{\acl{cte}} (\acs{cte})---in this application, it is distance between the 
aircraft nose wheel and the centerline of the taxiway or runway---together with 
the location of the nose wheel relative to the centerline. 
Effectively, the \ac{lec} performs regression under supervised learning to 
construct a mapping from a vector of real values (\ie images) to a signed, real valued scalar (\ie \ac{cte}). The absolute value gives the magnitude of the 
distance, and the sign indicates location, \ie to the left ($-$) or to the
right ($+$) of the centerline. 
%
%This is first estimated by the \ac{cnn} as the \emph{normalized} \ac{cte}: a 
%single real number in the interval $[-1, 1]$. The sign indicates the location, 
%and the value estimates the position, as a value proportional to the width of 
%the runway. Thus, in reference to the aircraft nose wheel, 
%%\begin{inparaenum}[\itshape i\upshape)] 
%%	\item 
%	\ac{cte} $= 0$ means that it is exactly on the runway centerline; 
%%	\item 
%	\ac{cte} $= -1$ means that it is to the extreme left of the runway 
%		centerline; and
%%	\item 
%	\ac{cte} $= +1$ means that it is to the extreme right of the runway 
%		centerline. 
%%\end{inparaenum}
%Similarly, $|$\ac{cte}$| = 0.03$, say, means that it is $2.25$\meter~offset 
%from the runway centerline, where the width of the runway $\approx45$\meter. 
%
%Other plausible and relevant output produced by processing the scenes can 
%include, for example, the \emph{heading error} (the angular distance between 
%the direction in which the aircraft nose is pointed and a reference direction, 
%usually that of the runway), the distance to different taxiway or runway 
%markings, 
%%---\eg \emph{side stripes} (designating taxiway or runway edges),  
%%\emph{aiming points} (visual references for landing), \etc---
%as well as detection and classification of objects in the path of the aircraft. 
%
%
%
The \ac{cnn} output of %normalized 
\ac{cte} is then input to 
%a Kalman filter that smooths and scales it (by the normalization 
%factor) to give a predicted \ac{cte} representing the actual distance between
%the nose wheel and the centerline. This is then fed to 
a traditional (\ie non-\ac{lec}) implementation of the navigation and 
control function that interprets the input and generates the appropriate 
signals, \eg thrust, steering, and flight control surface actuation (Fig.~\ref{f:lec-example}).

%\section{Application}\label{s:application}

%\subsection{Preliminaries}\label{ss:preliminaries}
%
%
%
%Next we identify the assurance concerns for the \ac{lec} of our running example
%(Section~\ref{s:example}, Fig.~\ref{f:lec-example}), after which we elaborate 
%the scope of system-level assurance.
%

\subsection{Scope of System-level Assurance}
\label{ss:system-level-assurance-scope}

The core functional objective for taxiing, whether under direct control of 
a human pilot or autonomously, is to be aligned with, and to safely follow, the 
taxiway or runway centerline when taxiing, during takeoff, and after 
landing~\cite{faa-flying-hdbk}. %\blurb{\eg avoid obstactles}

%either: 
%\begin{inparaenum}[\itshape i\upshape)] 
%	\item when taxiing from the parked position along a taxiway to the 
%	assigned runway for takeoff (\ie the taxi and takeoff flight phases); or 
%	\item when decelerating on the runway after touchdown, followed by taxiing 
%	on the taxiway until coming to a stop at the assigned parking position 
%	(\ie the landing flight phase). 
%\end{inparaenum}

In general, alignment with the centerline requires that the roll (longitudinal)
axis of the aircraft be parallel to the centerline and (possibly) offset on 
either side by no more than a specified threshold. %\blurb{$\pm 1.5$\meter}
From a system safety standpoint, among the main safety goals is to avoid 
\emph{runway excursions}, \ie veering off a runway or taxiway surface, either 
laterally, or by overshooting either end. 
%during the taxi, takeoff, and landing flight phases. 
%
For the autonomous taxiing capability in particular, these objectives are to be 
accomplished without pilot input, under a variety of diverse environmental 
situations, \eg nominal weather conditions, in low visibility, at night time, in 
rainy conditions, under crosswinds, \etc
%\footnote{However, a (remote or onboard) human pilot can intervene and assume 
%positive control over the aircraft for safety reasons: for example, after 
%processing any air traffic control instructions, or when there are obstacles 
%or ground traffic posing a collision hazard.}
%

%\blurb{Link the discussion in this section to assurance properties and assurance
%measures to show their applicability to system-level assurance}

For each of the objectives above, we can formulate system-level assurance 
properties linked to the dependability attributes of interest. 
%---chiefly safety, reliability, and correctness. 
Then, following our methodology 
(Section~\ref{ss:methodology})
%by decomposing and allocating the system-level objectives
we can specify the assurance properties/requirements) on the perception 
function, and on the \ac{lec} prediction of \ac{cte} in particular. 
These address both 
\begin{inparaenum}[(i)]
	\item the functional capabilities, \eg providing \ac{cte} as a signed 
	numerical value falling within a specified interval, and  
	\item the non-functional attributes, \eg providing reliable estimates of 
	\ac{cte} (considered next). %when there are input perturbations. 
\end{inparaenum}
%
%
%
%\blurb{Add a footnote explaining the relationship between reliability and 
%correctness}
%
%\blurb{Explain distinction between correctness/reliability at component level 
%and reliability at the system level}

\subsection{Assurance Properties and Assurance Measures}
\label{ss:assurance-concerns}

%For our running example of autonomous taxiing, we are interested in providing 
%assurance of the domain-specific use of the deep \ac{cnn} to compute \ac{cte}. 
%rather than domain-independent assurance properties \blurb{such as}
%and measures applicable for the deep \ac{cnn} itself (Fig.~\ref{f:lec-example})

We can further refine the objective of providing reliable estimates of \ac{cte}
into reliability assurance properties, including: 
\begin{inparaenum}[(i)]
	\item the sensor is operating and functional when required,
	\item \label{normal-op-sensor} a normally operating sensor produces an 
	accurate, precise, stable, and robust estimate\footnote{\rev{The 
	quantities of accuracy, precision, stability, and robustness are, in fact, 
	reliability assurance properties that can apply to distance/position 
	sensors in general.}} of 
	\ac{cte}, and
	\item \label{inoperable-sensor} an inoperable sensor either produces no 
	output, or produces an explicit error output. 
%	\item an inoperable sensor does not produce spurious output, \ie an output 
%	that could be misinterpreted as a legitimate estimate of \ac{cte}.
\end{inparaenum}
%
%Items \ref{normal-op-sensor}) and \ref{inoperable-sensor}) 
The latter two above together imply that there are no spurious outputs. 
Note that this is not a comprehensive enumeration of the applicable assurance
properties for the \ac{lec} of Fig.~\ref{f:lec-example}, but a representative 
example.  
The \ac{lec} in our running example can be considered to be a (software) 
sensor producing a measurement---in fact, an inference~\cite{Estler1999}---on 
position and location. Henceforth, we will refer to the \ac{lec} of the 
running example as the \emph{sensor}.

%We adapt the definitions that follow from~\cite{JCGM2012}.
\rev{There are generic, measurement-based definitions for accuracy, precision, stability, and robustness~\cite{JCGM2012}, which we now specialize for the sensor.}
%
%In addition to the assurance properties obtained from the decomposition 
%and refinement of system-level assurance properties (as discussed earlier 
%in Section~\ref{ss:system-level-assurance-scope}), we also adopt the assurance 
%properties applicable to position sensors as those relevant for the \ac{lec}: 
%\begin{compactitem}[--]
%\item 
\emph{Accuracy} is the \emph{closeness} of the \ac{lec} estimate of \ac{cte} to 
a reference (true) value (also known as the \emph{ground truth}) evaluated over 
all inputs. Accuracy is inversely proportional to the \emph{sensor 
error}: the difference between the estimated \ac{cte} and the true value that 
is known during training and in validation, but may not be known in testing 
and operation. The greater the accuracy, the lower the sensor error. 
%%
%%Since the sign of the \ac{cte} indicates nose wheel location relative to 
%%the centerline, we use the sensor error as the \emph{regression 
%%error}\footnote{However, in general, and depending on the type 
%%of regression being performed, there are other formulations for regression 
%%error, \eg \emph{mean squared error} (MSE), \emph{mean absolute error} (MAE), 
%%\etc}. 
%%
%%\blurb{clarify it is different from the sensor error} 
%%
%
%%If the \ac{lec} output were to be a probability distribution (instead of 
%%a deterministic value, as in our running example), the sensor error 
%%would be computed as the difference between the true \ac{cte} and the expected 
%%value of the sensor estimate of \ac{cte}. 
%
%%\item
%	
\emph{Precision} is the amount of variability in repeated estimates of \ac{cte}, 
for a given input. In operation or testing, a \ac{cnn} that has been trained in 
an offline supervised learning scheme (as in our running example) has fixed 
parameter values. Thus, for a given image, the output is fixed (\ie it is deterministic).  
In other words, the sensor response is not expected to change when a given 
image is repeatedly supplied as input. As such, the quantity of precision 
does not have a meaningful interpretation for a single image. 
%
%\blurb{Maybe say measurement precision? }

%\item
\emph{Stability} represents the amount of drift in \ac{cte} estimates, 
%in repeated inferences 
for a sequence of samples. For the \ac{lec} of the 
running example, quantifying stability involves determining the variation in 
the sensor error over a sequence of images that represent the time series 
progression of the camera images recorded as the aircraft moves on the 
runway during taxiing, takeoff, or after landing. 
%
%\item
\emph{Robustness} gives the amount of variability in the estimates of \ac{cte} 
%(or equivalently, the sensor error) 
in response to abnormal inputs. 
%
%\end{compactitem}

Other types of assurance properties can include expressions involving 
thresholds applied to the quantities indicated above, \eg that accuracy 
exceeds some minimum level. 
The assurance measure for each of the above assurance properties 
%(Section~\ref{ss:assurance-concerns}) 
is the respective probability distribution. For example, consider \ac{lec} 
accuracy for our running example: a useful assurance measure for this 
component-level (reliability) assurance property is the \ac{pdf} 
of the sensor error (Fig.~\ref{f:assurance-measures}),  %\ie the distribution of the \ac{lec} estimate of \ac{cte}. 
%The expected value of this distribution gives the most likely numerical 
%value for sensor error, while the variance gives a characterization of the 
%uncertainty (and, in turn, the confidence, \ie assurance) in the same. 
%%
whose expected value is the most likely numerical value for sensor error, 
and whose variance gives a characterization of the uncertainty (and, in turn, 
the confidence, \ie assurance) in the same. 

\subsection{Quantifying \ac{lec} Assurance}
\label{ss:lec-assurance-q}

To illustrate \ac{lec} assurance quantification in the context of our running 
example, we mainly focus on \ac{lec} accuracy (equivalently, sensor error). 
As previously mentioned, \ac{lec} precision is not meaningful in this context. 
We do not address (assurance of) stability, robustness, or other \ac{lec} assurance properties in this paper, due to space constraints.

\subsubsection{Gaussian Process Modeling}\label{sss:gp-modeling}

%
%\blurb{The previous sections described key assurance properties of
%LECs, such as accuracy robustness, and stability, both instantaneous
%and aggregated over arbitrary durations. Given these properties we
%then need to decide how to down-select those properties that are
%relevant for a given application of an LEC within a system for a
%particular mission, and then specialize the properties, and
%characterize them as statistical formulas that can then be implemented
%in the form of monitors.}
%

%\blurb{Now we consider various statistical representations for 
%assurance properties, consider their pros and cons, and describe some
%experiments to assess their \blurb{what?}.
%}

%\blurb{Needs cleanup from here}
%
%To this end, we evaluate the \ac{rlec} response at a known set of 
%test points, input and response of the \ac{rlec} to obtain the 
%\ac{rlec} test point errors to build the training set of the GP, 
%$(X,E)$, in which $X$ is the vector of \ac{rlec} inputs and $E$ 
%is the vector of the corresponding response error. The next step 
%is building a random process, in the form of a Gaussian random 
%process. The joint probability distribution of the observed set 
%of \ac{rlec}'s error, $E$ and the prediction error, $E^{*}$ at 
%a given \ac{rlec} input is represented as

We model the assurance measure for sensor accuracy as a stochastic error 
surface, using Bayesian non-parametric \ac{uq} techniques, in particular a 
\acf{gp} regression model~\cite{gp-book}. 

\newcommand{\gpinput}{\bm{\mathrm{X}}}
\newcommand{\testinput}{\bm{\mathrm{X'}}}
\newcommand{\sensorerr}{\bm{\mathrm{E}}}
\newcommand{\gppred}{\bm{\mathrm{E'}}}

Let $[\gpinput, \sensorerr]$ be the set of training data points, where $\gpinput$
is a vector of sensor inputs, and $\sensorerr$ is the vector of corresponding
sensor errors.
%
%Let $\gpinput$ be the vector of training input samples to the \ac{gp} for which 
%$\sensorerr$ is the corresponding vector of sensor errors. 
%
Let $\gppred$ be the \ac{gp} prediction of sensor error for an input from the 
test data, $\testinput$; then the joint probability distribution of the
observed sensor errors and the \ac{gp} prediction of sensor error is a 
multivariate Gaussian distribution: %given as:
%
%\begin{equation}
%	\begin{bmatrix}
%		\bm{E} \\ 
%		E^*
%	\end{bmatrix}
%	\sim 
%	\mathcal{N}\left(0, \bm{K}\right)
%\end{equation}
%where 
%\begin{equation}
%\bm{K} = 
%	\begin{bmatrix}
%		K(X,X) & K(X,X^*) \\ 
%		K(X^*,X) & K(X^*,X^*)
%	\end{bmatrix}
%\end{equation}
%
\begin{equation}\label{eq:gp}
	\begin{bmatrix}
		\sensorerr \\ 
		\gppred
	\end{bmatrix}
	\sim 
	\mathcal{N} \left(0, 
	\begin{bmatrix}
		K(\gpinput,\gpinput) & K(\gpinput,\testinput) \\ 
		K(\testinput,\gpinput) & K(\testinput,\testinput)
	\end{bmatrix}
	\right)
\end{equation}

$K(A, B)$ is a \emph{covariance matrix}, whose elements are the values of 
a suitable \emph{covariance function}. The latter is evaluated at all pairs of 
the training and test data, $(\gpinput, \testinput)$, and models the 
correlation between those data. For the running example, we have used the 
\emph{squared exponential} covariance function~\cite{gp-book}, which models the 
intuition that the data that are \emph{closer} to each other are more 
correlated, while those that are farther away are less correlated. 
The posterior distribution of the \ac{gp} prediction of sensor error is given 
from Bayesian update, as:
\begin{IEEEeqnarray}{rrcll}\label{eq:gp-update}
	\gppred \mid \gpinput, \testinput, \sensorerr & &\;\; \sim \;\; 
				& \mathcal{N} & (K(\testinput, \gpinput) 
								K(\gpinput,\gpinput)^{-1}\sensorerr, \\
   				      & &   & & K(\testinput, \testinput) 
				      			 -K(\gpinput,\gpinput)^{-1} 
							      K(\gpinput,\testinput))\;\; \nonumber
\end{IEEEeqnarray}

Note that other covariance functions may be chosen~\cite{gp-book} \rev{to
explore how different \ac{gp} models perform, although that tradeoff is not in 
scope for this paper.} 

%we have not done so in this work.

\begin{figure}[t]
	\centering
	\includegraphics[scale=0.5]{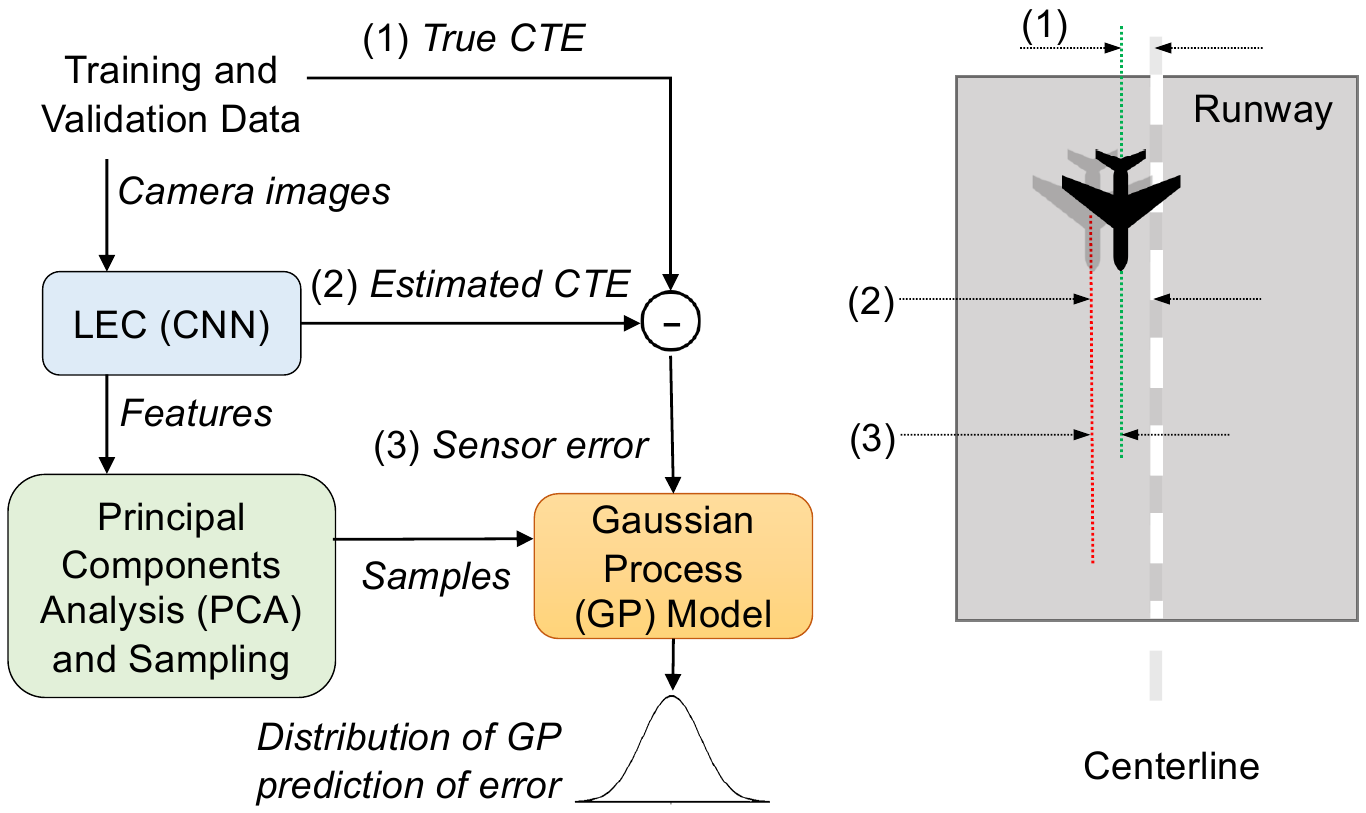}
	\caption{Training a \ac{gp} model to learn and predict sensor (\ac{lec}) 
	error.}
%	and the associated uncertainty.}
	\label{f:gp-model}
\end{figure}

Fig.~\ref{f:gp-model} shows how we build and train the \ac{gp} represented
by equations~\eqref{eq:gp} and~\eqref{eq:gp-update}.
We construct the training data for the \ac{gp} model from samples of the 
training and validation data of the \ac{lec}. More specifically, we use 
the features from the first fully connected layer after the convolutional 
layers, and apply \ac{pca} to reduce the dimensionality of the data set, 
$[\gpinput, \testinput]$. 
Then we apply clustering techniques---such as \emph{k-means clustering}---to 
select the samples supplied to the \ac{gp}. For these samples, since the true 
value of \ac{cte} is known a priori, we can also compute the sensor error, 
$\sensorerr$. 
%in the \ac{lec} estimate of \ac{cte}. 
For brevity, we omit other details of the \ac{gp} learning and 
training process, as well as the details on its computational implementation.

\subsubsection{Discussion}
\label{sss:gp-modeling-for-dummies}

The idea is to model (learn) the sensor error performance and extrapolate the results to predict sensor error ($\gppred$) in operation and testing. The extent 
to which the \ac{gp} is trained induces uncertainty in its prediction
of sensor error. This uncertainty is captured by the \ac{gp} output: a Gaussian 
distribution whose expected value is the prediction of sensor error that is 
most likely, and whose variance is the uncertainty in that prediction. 
This distribution captures the full scope of uncertainty in the 
predicted sensor error for a single sample input image, \ie the 
instantaneous assurance in \ac{lec} accuracy. 
Also, the posterior distribution is conditioned on \emph{all} data 
on which it has been trained, including the input on which the prediction 
is made. This is in contrast to the Bayesian parametric approach, where the 
posterior distribution of error is conditioned \emph{only} on the sensor error.

%
%Thus, for a sequence of samples, the \ac{gp} gives similar distributions, 
%which may or may not have the same expected value and variance. 
%
When the assurance property involves bounds on accuracy, \ie 
that the sensor error falls within a certain range, or that it does not exceed 
some required maximum value, then by evaluating the \ac{cdf} of the 
%predicted sensor error distribution 
\ac{gp} output at the bounds, we can compute the probability that 
the assurance property holds. In fact, since it is a posterior distribution 
computed from Bayesian update, the bounds represent the Bayesian 
\emph{credible interval} and the probability represents \emph{confidence}.

%The idea is to model (learn) the error performance of the sensor and 
%extrapolate the results to predict sensor error in operation and 
%testing. However, building a model using a limited number of training 
%points imposes uncertainty to the predicted LEC error, which is captured 
%by the GP response. Therefore, having defined the component assurance as the probability of maintaining LEC accuracy (denoted by LEC error) to be less than the required error constrain can be obtained by the CDF of the error distribution evaluated at the error constrains, given the error probability distribution 
%which is characterized by the GP outputs.
% 
%However, our problem does not require any constrain on the value of the error. 
%We therefore, simply capture the uncertainty of the predicted error through its corresponding probability distribution which represent the uncertainty of the predicted assurance property (accuracy) of the LEC.

%
The rationale underlying our approach %to \ac{uq} 
is that, as long as the \ac{gp} model is itself shown to be accurate 
relative to the sensor: 
\begin{compactitem}[--]
	\item For the \ac{gp} training inputs, the uncertainty in the predicted 
	error is zero and there is full confidence in the sensor performance, 
	\ie the expected value of the \ac{gp} output is the observed sensor 
	(\ac{lec}) error, and the variance in this expectation is zero.
	
	\item For inputs drawn from the \emph{training environment}---distinct from 
	the \ac{gp} training inputs---we can quantify the uncertainty in the 
	predicted sensor error (which is governed by the covariance matrix/functions
	chosen). In turn, when the bounds on accuracy are known, we can gauge the 
	confidence (assurance) in sensor accuracy as a probability. Note that, for 
	our running example, we did not have an accuracy requirement, hence we 
	characterize uncertainty, by providing the distribution of the predicted 
	sensor error. %rather than confidence. %as a probability value. 
	
	\item On inputs outside the training environment, there can be an abnormal 
	increase in the \ac{gp} output uncertainty, where a decision can be made as 
	to whether or not to trust the sensor (or even the \ac{gp} model for 
	that matter). Effectively, this can be a candidate runtime monitor for 
	detecting dataset shift. 
\end{compactitem}

\begin{figure*}
\centering
    \begin{subfigure}{0.3075\textwidth}
    	\centering
    	\includegraphics[width=\textwidth]{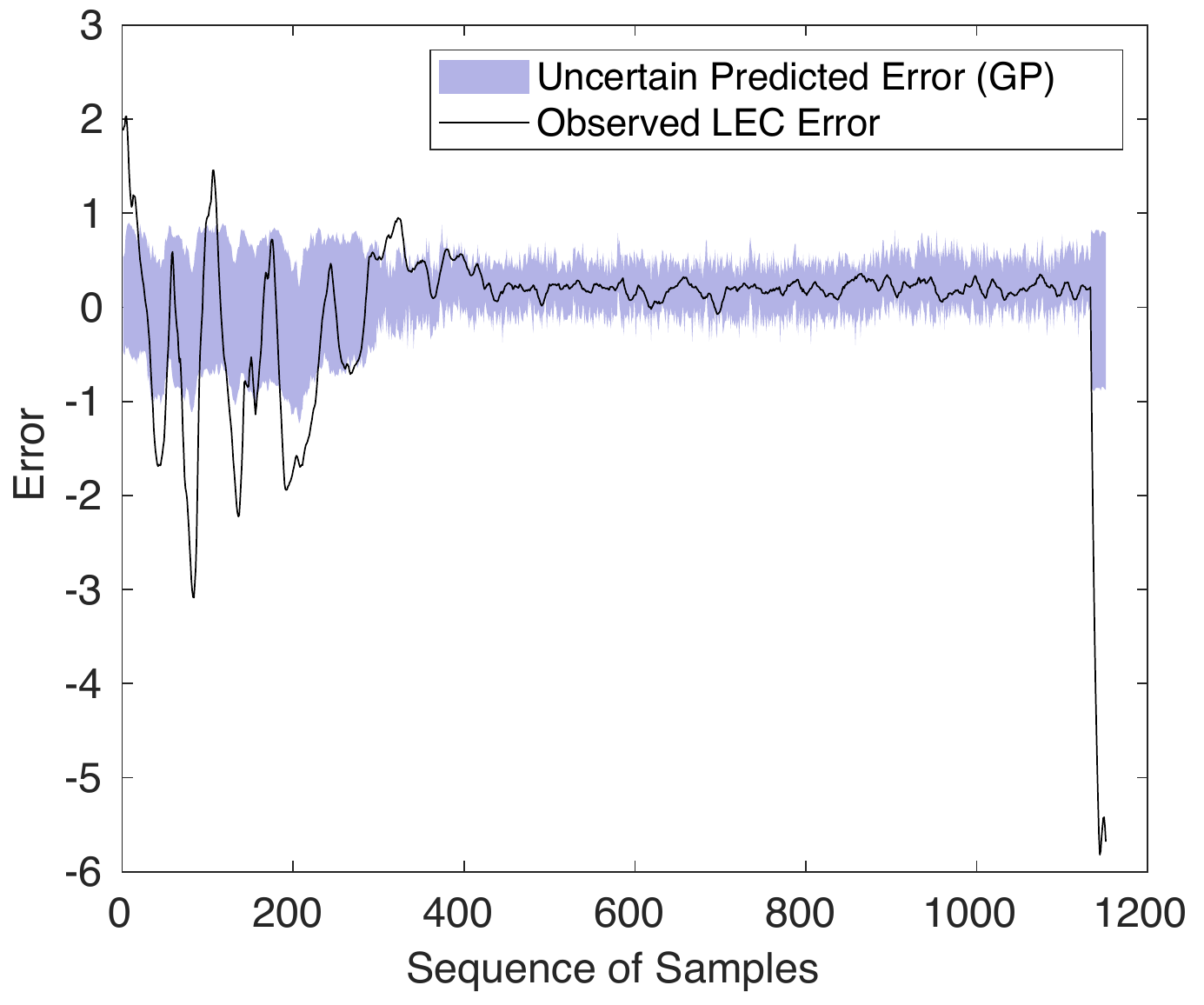}
    	\caption{}
    	\label{f:gp-error-prediction}
    \end{subfigure}
    \hfill
    \begin{subfigure}{0.3075\textwidth}
    	\centering
    	\includegraphics[width=\textwidth]{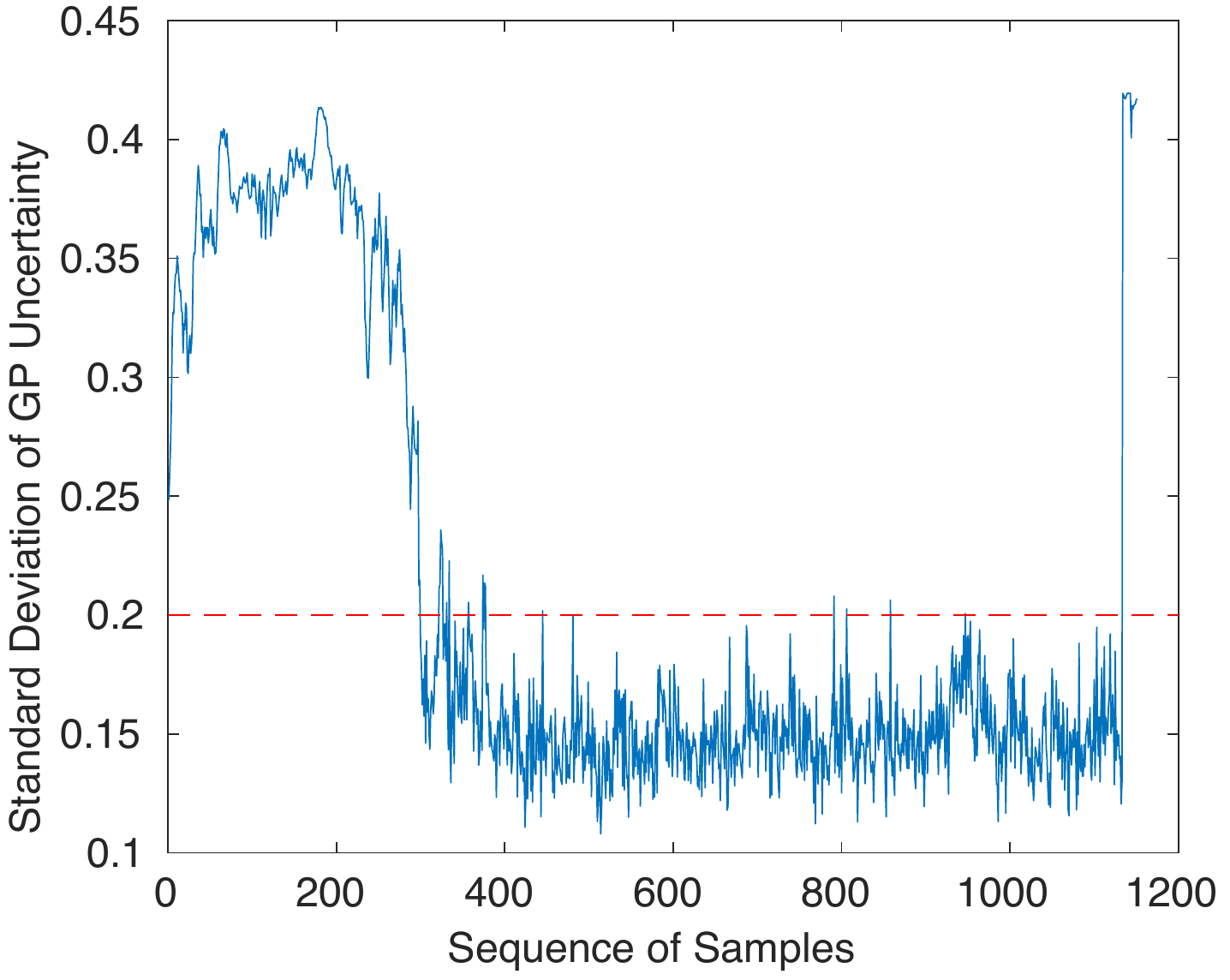}
    	\caption{}
    	\label{f:gp-uncertainty-std}
    \end{subfigure}
    \hfill
	\begin{subfigure}{0.3075\textwidth}
	\centering
	\includegraphics[width=\textwidth]{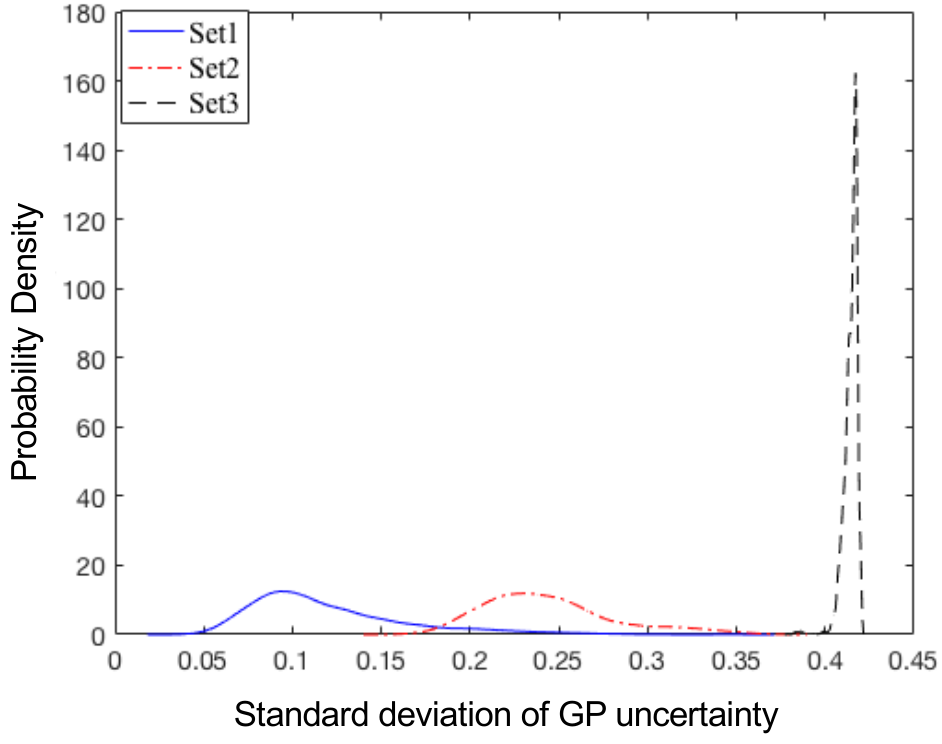}
	\caption{}
	\label{f:data-set-shift}
	\end{subfigure}
	\caption{Experimental results for quantifying \ac{lec} accuracy and assurance:
		(\subref{f:gp-error-prediction})
     	\ac{gp} uncertainty in the predicted sensor error for unseen (test) 
    	data drawn from the \setone data set. For the first $\approx 350$ samples, 
    	the \ac{gp} has a large uncertainty, \ie neither the \ac{gp} 
    	nor the \ac{lec} can be trusted. For the remaining samples, the 
    	\ac{gp} accurately predicts \ac{lec} error and the
		associated uncertainty
    	(\subref{f:gp-uncertainty-std}) 
    	Standard deviation of \ac{gp} uncertainty in the predicted sensor error, 
		for the same sequence of samples as in Fig.~\ref{f:gp-error-prediction}, 
		showing an abnormal increase in the uncertainty for the first $\approx 
		350$ samples
        (\subref{f:data-set-shift})
        Probability density of the standard deviation of \ac{gp} uncertainty in 
        the predicted sensor error, for the \setone, \settwo and \setthree data 
        sets. The density for \setone represents the uncertainty in the 
        predicted sensor error in the training environment; the densities 
        for \settwo and \setthree are noticeably shifted, indicating both 
        dataset shift and increased uncertainty in the \ac{gp} prediction.}
%        of sensor error.}
    \label{f:experimental-results}
\end{figure*}

\subsection{Experimental Results}\label{ss:experimental-results}

We now present some results of our experiments in quantitatively characterizing
\ac{lec} assurance (\ie \ac{lec} accuracy given in terms of the sensor error), 
%for the running example system, 
based upon data generated from a simulation platform.
%
%\blurb{Explain how the GP prediction allows us to determine whether the 
%LEC accuracy is acceptable. For this we need a threshold on what constitutes 
%sufficient accuracy, \ie how much error in the LEC estimate of \ac{cte} are 
%we willing to tolerate? say, 1/8th the width of the safety margin?}
%
In brief, we use a commercial-off-the-shelf 
flight simulator 
%\footnote{X-Plane Flight Simulator: 
%\href{https://www.x-plane.com/}{https://www.x-plane.com/}} 
instrumented to reflect the pipeline architecture of Fig.~\ref{f:lec-example}. 
The simulation
environment includes various airports from which we chose a specific airport  
%in which to simulate the overall aircraft system. 
having runway and taxiway surfaces with centerlines of varying quality (\eg 
portions of the centerline may be obscured at various locations on the 
runways). 
%
%The simulator additionally allows us to change various aspects of the 
%simulation, from which we have selected the following, as variation 
%points to create different training and test environments: 
%

The simulator additionally allows changing various aspects of the 
simulation (to create different training and test environments),
two of which we selected: 
\begin{inparaenum}[(i)]
%	\item The initial aircraft position before taxiing: at the runway threshold, 
%	elsewhere on the runway, nose aligned with the runway, nose not aligned 
%	with the runway. 
	\item weather conditions: \emph{Clear}, \emph{Overcast}, \emph{Foggy}, 
	\emph{Low visibility}, and \emph{Scattered clouds}; and 
	\item the time of day: \emph{7am}, \emph{Noon}, \emph{2pm}, and 
	\emph{6pm}. 
\end{inparaenum} 	
For example, some environments 
%(given the initial aircraft position as located at the runway threshold, 
%and having the nose aligned with the runway), 
are: 
\emph{Clear Noon}, \emph{Overcast Noon}, \emph{Foggy Noon}, \emph{Clear 6pm}, 
\emph{Foggy 7am}, \etc
For these, we gathered images via automated screen-capture 
(simulating the camera output) whilst taxiing the aircraft on the airport 
runway, not only using different software controllers, but also under manual 
control (via an external joystick and throttle). In parallel, for each image, 
we gathered actual \ac{cte} (from internal simulation variables), and \ac{lec} 
estimates of \ac{cte} (henceforth, referred to as \emph{sensor outputs}). 

The combination of images, corresponding actual \ac{cte} values (serving as 
ground truth), and sensor outputs comprise a data set. We used several data sets, 
one for each of the different environments identified above. Of these,
samples drawn from the \emph{Clear Noon} data set were used to train 
the \ac{lec}. That is, the \emph{\ac{lec} training data} 
comprises %\blurb{how many?} 
samples of the pair of input image and corresponding actual \ac{cte}, drawn from 
the data set representing the \emph{Clear Noon} training environment. 

Then, we ran the trained \ac{lec} on $5600$ samples drawn from data sets 
representing the \emph{Overcast Noon} test environment as well as the \emph{Clear Noon} training environment, to obtain the sensor output (and from it, the 
sensor error). From this collection of data, we drew $600$ samples (as described earlier in Section~\ref{sss:gp-modeling}, and Fig.~\ref{f:gp-model})
to serve as the \emph{\ac{gp} training data}. In other words, the \ac{gp} 
training data comprises samples from the \ac{lec} training \emph{and} \ac{lec} 
test environments. 
%Note that \ac{gp} training data are not randomly drawn images, 
Finally, to test the \ac{gp} we used the following data sets, samples from which 
were unseen by the \ac{gp}: 
\setone: \emph{Clear Noon} and \emph{Overcast Noon} (excluding the $600$ 
samples that are the \ac{gp} training data); \settwo: \emph{Foggy Noon}; 
and \setthree: \emph{Clear 6pm}.

%\begin{figure}
%	\centering
%	\includegraphics[width=0.72\columnwidth]{error-prediction}
%	\caption{\ac{gp} uncertainty around the sensor error for unseen (test)
%	data drawn from the (Set1) data set. %, \ie 
%	%the difference in the \ac{lec} estimate of \ac{cte} and the 
%	%true \ac{cte}. 
%	For the first $\approx 400$ samples, the \ac{gp} has a large 
%	uncertainty and effectively neither the \ac{gp} nor the \ac{lec} can 
%	be trusted. For the remaining samples, the \ac{gp} predicts the error
%	performance of the \ac{lec} and supplies the associated uncertainty.}
%	\label{f:gp-error-prediction}
%\end{figure}
%
%\begin{figure}
%	\centering
%	\includegraphics[width=0.75\columnwidth]{gp-uncertainty-std}
%	\caption{Standard deviation of the uncertainty in the \ac{gp} prediction
%	of sensor error, plotted for the same sequence of samples shown in 
%	Fig.~\ref{f:gp-error-prediction}, where an abnormal increase in the 
%	uncertainty can be seen for the first $\approx 400$ samples.  }
%	\label{f:gp-uncertainty-std}
%\end{figure}
%
%\begin{figure}
%	\centering
%	\includegraphics[width=.75\columnwidth]{data-set-shift}
%	\caption{Uncertainty in the \ac{gp} prediction of sensor error
%	represented as a probability density of the standard deviation for 
%	the (Set1), (Set2) and (Set3) data sets.}
%	\label{fig:std-error-combined}
%\end{figure}

For all data sets in our experimental setup, because we know the true \ac{cte}, 
we can compute the actual sensor error. %, as the difference
%of ground truth and the sensor output, \ie \ac{lec} estimate of \ac{cte}. 
%
Also, intuitively, since we expect the \ac{gp} to perform well (accurately 
predict sensor error) in its training environment, to characterize 
the \emph{quality} of the \ac{gp} we compare its prediction of sensor 
error against the actual sensor error for the data samples from \setone.  
Since the \ac{gp} output is a Gaussian distribution (with different mean and variance for each test input), we determine the fraction of all 
%frequency (as a percentage) 
sensor errors from \setone that fall within $1\sigma$ and $2\sigma$ around the 
mean value of the \ac{gp} output.
% Gaussian distribution of the error predicted by GP, 
These are, respectively, $68\%$ and $91\%$. 
Note that both $\sigma$ (standard deviation) and the mean value are predicted 
as parameters of the \ac{gp} output. Since the values within $1\sigma$ and 
$2\sigma$ around the mean value account for about $68\%$ and $95\%$ of 
normally distributed data, the \ac{gp} can be considered to be reasonably 
accurate in its training environment. 

Fig.~\ref{f:gp-error-prediction} visually illustrates this, plotting 
\begin{inparaenum}[(i)]
	\item the area within $2\sigma$ around the expected value of the \ac{gp}
	prediction of sensor error, against
	\item the actual value of sensor error,
\end{inparaenum}
for a \emph{Clear Noon} data set, where the aircraft was controlled by a 
software controller. From the figure, we see that the \ac{gp} accurately 
predicts \ac{lec} error for the sequence of samples $350$--$1200$. 
We additionally observe (from Figs.~\ref{f:gp-error-prediction} 
and~\ref{f:gp-uncertainty-std}) that for the first $\approx 350$ samples, 
the \ac{gp} uncertainty in the predicted sensor error is not only large, 
but also the $2\sigma$ bound does not include the actual sensor error for 
some samples. 

To diagnose this observation, we plot (in Fig.~\ref{f:data-set-shift}) the 
distribution of the standard deviation of the %\ac{gp} prediction of sensor error
%(\ie 
the \ac{gp} uncertainty in its prediction of sensor error, for all three
data sets, \ie the unseen test set (\setone) drawn from the \ac{gp} training 
environment; and the tests sets from the unseen environments, \settwo and \setthree.
Fig.~\ref{f:data-set-shift} indicates that the distributions are, evidently, 
different. We can view this as an indicator of data distribution shift. 
That is, in the unseen environments, neither the \ac{gp} nor the \ac{lec} can 
be trusted. 
Since the distribution of \setone is from the training environment where
the \ac{gp} is known to perform reasonably well, we show the $2\sigma$ bound of 
that distribution (as shown by the dotted red line in 
Fig.~\ref{f:gp-uncertainty-std}) to discriminate an abnormal increase in 
uncertainty. As such (in Fig.~\ref{f:gp-uncertainty-std}), for the first 
$\approx 350$ samples from the \setone data set, we hypothesize that the 
increase in uncertainty could be attributed to
%\begin{inparaenum}[\itshape i\upshape)]
%	\item the likely presence of features in the data that are similar to 
%	data drawn from unseen environments, on which the \ac{gp} was not trained, or
%	\item 
	insufficient sampling of the training environment whilst creating the 
	\ac{gp} training data.
%\end{inparaenum}
%
This simple indicator of dataset shift deserves a more in-depth
study, \rev{which we will investigate as future work}. %that is not in scope for this paper. 

\section{Concluding Remarks}\label{s:conclusions}

%\paragraph{Utility, Generalizability, Challenges, and Future Potential}
%\label{ss:pros-cons}
%\paragraph{Reflection}

So far as we are aware, this work is the first to connect dependability 
attributes, \ac{uq}, and their applicability to \ac{lec} assurance. 
Conservatively, we expect that our approach is generalizable to those \acp{lec} 
trained in an offline, supervised learning scheme, addressing regression and 
classification problems where there is a reference or \emph{true} value 
against which \ac{lec} output can be compared. 
%
%In its current form, however, our methodology largely applies to \acp{lec} 
%that have been trained in an offline supervised learning scheme. In
%particular, we believe it can be applied to regression and classification 
%problems, where there is a notion of reference values against which \ac{lec} 
%output can be compared. 
%
%There are challenges to such generalizability, which 
%we discuss subsequently in this section. 
%
In this paper, we have shown the applicability to a regression problem 
using a worked aviation-related example. 
Exploring how our approach applies to \acf{rl}, is an avenue 
for further research. 

By using \acp{gp} for \ac{uq}, our approach implicitly assumes that the 
training and test data belong to a joint multivariate Gaussian distribution. 
Additionally, we have assumed a specific form for the covariance function
(squared-exponential) in our example, although others can be chosen.
\rev{Similarly, we can choose other non-parametric priors instead of \acp{gp}. 
In this paper, we have neither explored the suitability nor compared the 
performance of using alternative covariance functions or non-parametric priors. 
This---and more generally, comparing the suitability of different \ac{uq} 
techniques for assurance quantification---is another line of future work}. 

%\acp{gp} offer attractive properties besides a non-parametric representation, 
%such as analytical forms for the marginal and predictive distributions, 
%covariance functions facilitating model selection, and the availability of 
%tools and algorithms enabling computational implementation. However, there are 
%alternative non-parametric priors for \ac{uq}, though here we have neither 
%explored their suitability nor compared their performance. This---and more 
%generally, comparing the suitability of different \ac{uq} techniques for 
%assurance quantification---is another line of future work. 
%

%since they are not constrained to a particular 
%functional form, 
%admitting a potentially infinite parameter space. 
%
%For example, the scope of uncertainty being modeled can be expanded to include 
%the uncertainty in the \ac{gp} model parameters themselves, although determining 
%the hyper-parameters will require Monte Carlo sampling of the random process. 

%\blurb{Potential limitations - the GP is not trained directly on the images on
%which the LEC is trained; rather on the features extracted by the LEC. Thus any
%problems that have occurred during feature extraction are propagated to the GP,
%\eg adversarial perturbations. There is additionally a PCA performed on the 
%features, and if some features that have a large impact are excluded, then the 
%GP performance may be affected. Moreover, the samples are selected by k-means 
%clustering, but this is known to be inefficient.}
%
%As indicated in Section~\ref{ss:methodology}, 
\acp{gp}, and Bayesian non-parametric \ac{uq} approaches are generally more 
flexible than Bayesian parametric techniques, but command a larger computational
cost.
%
%However, this flexibility usually commands a larger computational 
%cost. 
%and may generalize less easily than a solution that uses Bayesian parametric \ac{uq}. 
\acp{gp} are known to scale poorly to high-dimensional data. 
Indeed, recall that we train the \ac{gp} on the features that 
the \ac{lec} extracts (Section~\ref{ss:lec-assurance-q}). 
%(rather than on the images used 
%for \ac{lec} training), which we then follow with a \acf{pca} (to further 
%reduce dimensionality) and clustering (to select \ac{gp} training samples). 
%
Thus, problems affecting feature extraction---\eg feature similarity for 
markedly different images/inputs---and the possibility of the \ac{pca} 
excluding features that potentially have a large impact, can affect \ac{gp} 
training and, thereby, its predictive performance. Similar to \acp{lec}, 
the selection of the \ac{gp} training samples also affects its predictive 
performance. A third avenue for future work is, therefore, exploring techniques 
to address both the dimensionality problem of \acp{gp}, and the selection of 
\ac{gp} training data. Promising results about the equivalence between 
\acfp{dnn} and \acp{gp}~\cite{lee2018}, and so-called \emph{sparse} 
\acp{gp}~\cite{snelson2006}, may provide a way forward.
\rev{Another candidate line of future work is to investigate how potential alternative approaches to \ac{uq} (such as Bayesian NNs, and training \acp{lec} with \emph{dropout}~\cite{gal2018}, which quantify uncertainty in 
the \ac{lec} output and not directly in assurance measures) may augment 
our proposed approach, \eg through \emph{ensemble} techniques.}
%

%
%Potential candidates to quantify the uncertainty of the actual value predicted by LEC, and therefore the LEC error, are Bayesian NNs, training the LECs with dropout [21] or ensemble of uncertainty quantification methods, where the expected value of error obtained from the first two is zero. However, in this study we constrained ourselves not to change the structure of the given LEC or retrain it. We also note that, comparing different means of uncertainty quantification techniques is out of the scope of this paper.

%
%We note that although the focus of this work has been \ac{lec} assurance, our 
%broader goal is to quantify the assurance of the wider \acf{les} that contains 
%the \acp{lec}. 
%
Although component- and system-level assurance can be related 
through a chain of linked assurance properties and measures---\eg
in an assurance argument---relating the quantification approaches is 
an important avenue of future work. 
Put in the context of our running example, this will seek, for instance, 
to establish how \ac{gp} uncertainty in the predicted sensor error relates 
to system-level assurance, given as the probability that the actual \ac{cte} 
does not exceed a required lateral offset threshold during taxi operations.

We have put forth a notion of assurance for \acfp{lec} based on identifying and
relating the relevant dependability attributes to application-specific 
assurance properties and measures. Our approach to assurance quantification is 
to give a probabilistic characterization of assurance measures, towards 
capturing uncertainty (and, thereby, confidence) \eg using Bayesian non-parametric techniques for \acl{uq}. To illustrate (and practically ground) our approach, we have applied it to an aviation system example. In particular, using \ac{gp} 
regression, we have quantified a specific assurance property (accuracy) for 
an \ac{lec} in its use as a sensor.
We have also presented the results from simulation experiments, discussing 
the assumptions made, and the challenges that remain.
This paper has focused on \ac{lec} assurance at the component-level. Ultimately, 
however, \emph{system-level} assurance---of the \acf{les} 
that contains the \acp{lec}---is what matters.
%
%Our focus here has been at the \emph{component-level} but, ultimately, 
%\emph{system-level} assurance is what matters. 
%
We are currently developing techniques to relate the two, as part of a wider \acf{dac} framework that will incorporate other sources of assurance (\eg 
classical runtime verification) towards a broader goal of dynamic assurance and, eventually, certification.

\bibliographystyle{IEEEtran}

%\bibliography{adp-edcc2019}


\end{document}